\documentclass{article} 
\usepackage{iclr2021_conference,times}

\usepackage{amsmath,amsfonts,bm}

\def\eqref#1{equation~\ref{#1}}

\def\1{\bm{1}}

\DeclareMathAlphabet{\mathsfit}{\encodingdefault}{\sfdefault}{m}{sl}
\SetMathAlphabet{\mathsfit}{bold}{\encodingdefault}{\sfdefault}{bx}{n}

\usepackage{hyperref}
\usepackage{url}
\usepackage{graphicx}

\title{Model discovery in the sparse sampling regime}
\iclrfinalcopy

\author{Gert-Jan Both, Georges Tod \& Remy Kusters  \\
Université de Paris, INSERM U1284 \\
Center for Research and Interdisciplinarity (CRI) \\
F-75006 Paris, France \\
\texttt{remy.kusters@cri-paris.org} \\
}

\begin{document}

\maketitle

\begin{abstract}

To improve the physical understanding and the predictions of complex dynamic systems, such as ocean dynamics and weather predictions, it is of paramount interest to identify interpretable models from coarsely and off-grid sampled observations. In this work we investigate how deep learning can improve model discovery of partial differential equations when the spacing between sensors is large and the samples are not placed on a grid. We show how leveraging physics informed neural network interpolation and automatic differentiation, allow to better fit the data and its spatiotemporal derivatives,  compared to more classic spline interpolation and numerical differentiation techniques. As a result, deep learning based model discovery allows to recover the underlying equations, even when sensors are placed further apart than the data’s characteristic length scale and in the presence of high noise levels. We illustrate our claims on both synthetic and experimental data sets where combinations of physical processes such as (non)-linear advection, reaction and diffusion are correctly identified.

\end{abstract}

\section{Introduction}
Mathematical models are central in modelling complex dynamical processes such as climate change, the spread of an epidemic or to design aircrafts. To derive such models, conservation laws, physical principles and phenomenological behaviors are key. However, some systems are too complex to model with a purely bottom up approach \cite{bolton2019applications,sanchez2020artificial}. When observational data is present, automated model discovery tools are becoming increasingly more useful to derive partial differential equations directly from the data. The classical method for data driven model discovery is to apply sparse regression on a set of pre-selected features, the so-called library. In the case of partial differential equations, this library is constructed from a set of (higher)-order spatial derivatives. Model discovery is thus effectively a two-step process: first construct the library, then apply sparse regression. Numerically differentiating the data to construct the library using finite differences is extremely sensitive to noise; in practice, usually splines are fitted first and then differentiated. Splines model the data as piece-wise polynomials, but this expansion breaks down when the spacing between two sensors is large. These methods, which we refer to as classical methods, thus fundamentally limit model discovery to densely sampled data sets: even when no noise is present, the error incurred by the numerical differentiation corrupts the library and renders the sparse regression algorithm useless. The limits of classical interpolation methods have long been known and are often cited as a reason to use neural networks instead. Automatic differentiation can then be used to calculate the derivatives \cite{baydin2017automatic}, resulting in much more accurate derivatives. Previous works \cite{long2018,both2019,both2020} have shown that using neural networks to create a surrogate of the data allows model discovery in noisy and small data sets. 

In this paper we systematically study how sample spacing influences model discovery and compare neural-network based interpolation with classical methods. Our focus is the influence of the differentiation method used to construct the (higher-order) derivatives and its impact on model discovery, in particular when the spacing between two sensors $\Delta x$ is larger than the underlying equations' characteristic length scale $l_c$. As NN-based model discovery method we use DeepMoD, which is able to combine NN-based interpolation with any sparse regression method \cite{both2019,both2020}. By using an identical sparse regression algorithm for both the classical method and DeepMoD, we can isolate the effect of interpolation on the library and the discovered equation. Our results show that NN-based interpolators, in contrast to classical methods, can recover the underlying equation when $\Delta x > l_c$. Furthermore, we show that NN-based interpolation can succeed even when $\Delta x \gg l_c$ by either randomly sampling or displacing the sampling grid over time. We corroborate our findings with experimental data sets of the 2D advection-diffusion equation and the 1D cable equation. In both cases, DeepMoD, discovers the underlying equation in this sparsely sampled regime, contrarily to classical methods. Our findings solidify the case for deep learning methods by showing that they succeed in a regime where classical methods fundamentally fail.

\section{Related works}

\paragraph{Sensor placement} 
There exists a vast literature on determining optimal sensor placement for control theory or signal reconstruction based on a library of features, emphasizing the importance of sampling in the limit of sparse data \cite{brunton2013,manohar2018,wang2019}. While many of these sampling strategies have been developed to either reconstruct multi-scale data-sets \cite{champion2019}, flow-fields \cite{brunton2015,loiseau2017sparse} or other physical properties of a system \cite{schaeffer2018extracting}, research on the exact role of spatial and temporal sensor density or distribution for model discovery has received limited attention. 

\paragraph{Sparse regression-based model discovery} Using sparse regression to discover differential equations was popularized by algorithms such as SINDY \cite{brunton_discovering_2016} and PDE-find \cite{rudy_data-driven_2017} and has received considerable interest for both ODEs \cite{mangan2017,messenger2020} as well as for PDEs \cite{rudy2017,long2018, vaddireddy2020}. These approaches have since been expanded to automated hyper-parameter tuning \cite{champion_data-driven_2019, maddu_stability_2019}; a Bayesian approach for model discovery using Sparse Bayesian Learning \cite{yuan_machine_2019}, model discovery for parametric differential equations \cite{rudy_deep_2019}.

\paragraph{Deep learning-based model discovery} With the advent of Physics Informed Neural Networks \cite{raissi_physics_2017, raissi_physics_2017-1}, a neural network has become one of the prime approaches to create a surrogate of the data and perform sparse regression either on the networks prediction \cite{schaeffer_learning_2017, berg_data-driven_2019} or within the loss function of the neural network \cite{both2019,both2020}. Alternatively, Neural ODEs were also used to discover the unknown governing equation \cite{rackauckas_universal_2020} from physical data-sets. Different optimisation strategy based on the method of alternating direction is considered in \cite{chen_deep_2020}, and graph based approaches have been developed recently \cite{seo_differentiable_2019, sanchez-gonzalez_graph_2018}. Finally, \cite{cranmer_lagrangian_2020, greydanus_hamiltonian_2019} directly encode symmetries in neural networks using respectively the Hamiltonian and Lagrangian framework.

\section{Methods}

\paragraph{Sparse regression}

A popular approach to discover a PDE from a spatio-temporal data set is to apply sparse regression on a library of candidate terms $\Theta$, e.g. solve, 
\begin{equation}
    u_t = f(1,u, u_x, ...) =   \Theta \cdot \xi,
\end{equation} 
to obtain the coefficient vector $\xi$. Here $u_t$ is the temporal derivative and each column in $\Theta$ is a candidate term for the underlying equation, typically a combination of polynomial and spatial derivative functions (e.g. $u$, $u_x$, $uu_x$). To promote the sparsity of this solution an $l_1$ regularization is added to the problem, leading to the so-called Lasso regression:

\begin{equation}
    \xi^* = \min_{\xi} \left \lVert u_t - \Theta \cdot \xi\right\rVert^2 + \lambda \sum_i |\xi_i|.
\label{eq:reg}
\end{equation}

Here $\lambda$ controls the strength of the regularization, and hence the resulting level of sparsity. In this paper we use the Lasso as a sparse regression algorithm, with $\lambda$ determined by 5-fold cross-validation. The resulting coefficients are normalized by the $l_2$ norm of the feature vector, $\hat{\xi}_i = \xi_i\cdot ||\Theta_i||_2 / ||u_t||_2$ and thresholded. The exact value of the threshold can significantly influence the resulting equation. We use exactly the same Lasso and threshold for both DeepMoD and the classical methods so as to eliminate the influence of the exact variable selection algorithm used. Our goal is to compare how the feature library $\Theta$ and temporal derivative $u_t$ as generated by either DeepMoD or a classical method differ, and its resulting effect on model discovery.

\paragraph{Numerical differentiation}

The features of the library $\Theta$ consists of (higher-order) derivatives, which need to be differentiated from the observed data. Numerical differentiation is typically performed either by finite differences or by fitting a spline on the data and subsequently differentiating this spline. Finite difference methods directly operate on the observed data to calculate the derivative. In this paper, we use a standard second order accurate central difference scheme. Finite differences is computationally cheap and easy to scale to higher dimensions, but suffers from sensitivity to noise and requires samples to be closely spaced for accurate results; the truncation error of the scheme given above scales with the grid sampling, $h$, as $\mathcal{O}\left( h^2\right)$. In the sparse regime where $\Delta x \to l_c$, higher order schemes will not further improve this method. Furthermore, FD requires samples on the edges of the domain to be discarded, and for small data-sets and higher order schemes this can become a significant fraction of the total data.

A more accurate and widely used alternative is to fit a spline to the data and differentiate it. When fitting using splines, the data is approximated by a piece-wise polynomial with enforced continuity at the edges. Splines yield more accurate results in practice, but do not scale easily to higher dimensions, especially when using splines of higher order. This hinders model discovery, which requires these higher orders due to the derivatives in the feature library; by using a fifth-order spline to approximate the data, we effectively approximate the 3rd order derivative with only a second order polynomial, hence hindering its application to model discovery.

\paragraph{DeepMoD} \cite{both2019,both2020}\footnote{github.com/PhIMaL/DeePyMoD} is a neural network-based model discovery algorithm. It uses a neural network to learn both a noiseless surrogate of the data $\hat{u}$ and a coefficient vector $\xi$ by minimizing,

\begin{equation}
\mathcal{L} = \frac{1}{N}\sum_{i=1}^{N}\left( u_i - \hat{u}_i \right) ^2 +\frac{1}{N}\sum_{i=1}^{N}\left( (\hat{u}_t)_i - \Theta_{i}(\xi \cdot g)\right)^2.
\label{eq:deepmod}
\end{equation}

Here $\Theta$ and $\hat{u}_t$ are calculated by automatic differentiation of the neural network output $\hat{u}$. $g$ is a mask which sets the active terms, i.e. the terms that feature in the differential equation. The first term learns the data mapping $(x, t) \to \hat{u}$, while the second term constrains the network to solutions of the partial differential equation given by $\hat{u}_t, \Theta$ and $\xi \cdot g$. During training, the coefficients $\xi$ are determined by solving the least squares problem corresponding to the second part of eq. \ref{eq:deepmod}. The mask $g$ is updated separately by a sparse regression algorithm. The mask $g$ thus selects which terms are in the equation, while $\xi$ are the coefficients of these active terms. The value of the threshold can impact the discovered equation. To remove this factor from our comparison, we use exactly the same method to find the sparse coefficient vector $\xi^*$ for DeepMoD and the classical methods. More details on DeepMoD can be found in \cite{both2019,both2020}.

We emphasize two key differences with classical methods: 1) DeepMoD uses automatic differentiation to calculate $\Theta$, and the accuracy of the derivatives is thus not fundamentally linked to the sample distancing as with numerical differentiation. 2) By including the regression term within the loss function, we regularise the solution of the neural network $\hat{u}$, with the learned solution of the PDE. The result of fitting a spline is solely based on the data, whereas with DeepMoD it is also influenced by the constrained of the underlying equation (i.e. $\xi$ and $g$). We show in the next section that these two differences allow model discovery in extremely sparse and noisy data sets, whereas classical methods fail.

\section{Results}

\subsection{Synthetic data - Burgers equation} 

We consider a synthetic data set of the Burgers equation $u_t = \nu u_{xx} - u u_x$, with a delta peak initial condition $u(x, t=0) = A \delta(x)$ and domain $t \in [0.1, 1.1], x \in [-3, 4]$. This problem can be solved analytically (see Appendix A) to yield a solution dependent on a dimensionless coordinate $z = x / \sqrt{4 \nu t}$. We recognize the denominator as a \textit{time-dependent} length scale: a Burgers data set sampled with spacing $\Delta x$ thus has a time-dependent dimensionless spacing $\Delta z(t)$. We are interested in the smallest characteristic length scale, which for this data set is $l_c = \sqrt{4 \nu t_0}$, where $t_0=0.1$ is the initial time of the data set. We set $A=1$ and $\nu=0.25$, giving $l_c = \sqrt{0.1} \approx 0.3$.

Splines do not scale effectively beyond a single dimension, making it hard to fairly compare across both the spatial and temporal dimensions. We thus study the effect of spacing only along the spatial axis and minimize the effect of discretization along the temporal axis by densely sampling 100 frames, i.e. $\Delta t = 0.01$. Along the spatial axis we vary the number of samples between 4 and 40, equivalent to $0.5 < \frac{\Delta x}{l_c} < 5$. We study the relative error $\epsilon$ as the sum of the relative errors for all the derivatives, normalized over every frame, 

\begin{equation}
    \epsilon = \sum_{i=1}^{3} \left\langle\frac{\left\lVert\partial_{x}^{i} u_j - \partial_{x}^i \hat{u}_j\right\rVert_2}{\left\lVert \partial_{x}^{i} u_j \right\rVert_2}\right\rangle_j
    \label{eq:error}
\end{equation}

where $i$ sums the derivatives and $j$ runs over the frames. The derivatives are normalised every frame by the $l_2$ norm of the ground truth to ensure $\epsilon$ is independent of the magnitude of $u$. $\epsilon$ does not take into account the nature of noise (e.g. if it is correlated and non-gaussian), nor if the correct equation is discovered. However, taken together with a success metric (i.e if the right equation was discovered), it does serve as a useful proxy to the quality of the interpolation. 

\begin{figure}
    \centering
    \includegraphics[width=0.9\textwidth]{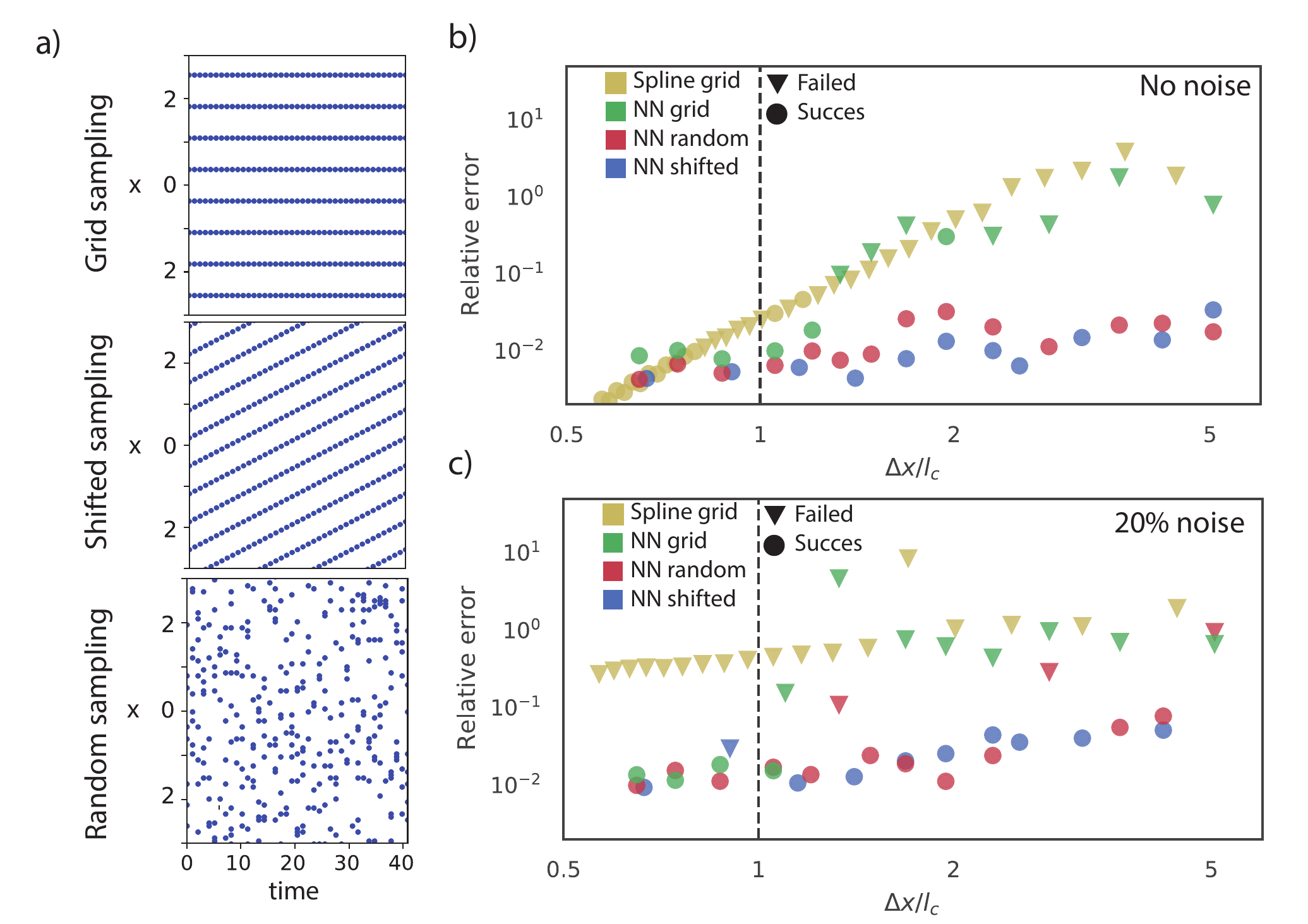}
    \caption{a) The three sampling strategies considered. b) and c) Error in the function library (Eq. \ref{eq:error}) as function of the distance between the senors $\Delta x$, normalized with $l_c = \sqrt{4 \nu t_0}$, for b) noise-less data and c) 20$\%$ of additive noise. The yellow symbols correspond to a spline interpolation and the green, blue and red correspond to the NN-based model discovery with various sampling strategies. The circles indicate that model discovery was successful while the triangles indicate that the incorrect model was discovered. The horizontal dashed line indicates the smallest characteristic length-scale in the problem: $\Delta x / l_c = 1$. }
    \label{fig:burgers}
\end{figure}

Figure \ref{fig:burgers}b) shows $\epsilon$ as a function of the relative spacing $\Delta x / l_c$ and whether the correct equation was discovered. The error when using splines (yellow) increases with $\Delta x$ and, as expected, we are unable to discover the correct equation for $\Delta x > 0.8 l_c$ (dots indicate the correct equation is discovered and triangles indicates it failed to do so). Considering the NN-based DeepMoD method, sampled on a grid (green), we observe that it is able to accurately interpolate and discover the correct equation up to $\Delta x \approx 1.2 l_c$.  The reason for this is that NN-based interpolation constructs a surrogate of the data, informed by both the spatial and the temporal dynamics of the data set, while classical interpolation is intrinsically limited to a single time frame. 

In figure \ref{fig:burgers}c) we consider the same graph with $20\%$ white noise on the data. Despite smoothing, the spline is unable to construct an accurate library and fails to discover the correct equation in every case. DeepMoD stands in stark contrast, discovering the correct equation with comparable relative error as in the $0\%$ noise case. 

\paragraph{Off-grid sampling} Whereas higher-order splines are constrained to interpolating along a single dimension, DeepMoD uses a neural network to interpolate along both the spatial and temporal axis. This releases us from the constraint of on-grid sampling, and we exploit this by constructing an alternative sampling method. We observe that for a given number of samples $n$, DeepMoD is able to interpolate much more accurately if these samples are randomly drawn from the sampling domain. We show in figure \ref{fig:burgers}b and c (Red) that the relative error $\epsilon$ in the sparse regime, can be as much as two orders of magnitude lower compared to the grid-sampled results at the same number of samples. We hypothesize that this is due to the spatio-temporal interpolation of the network. By interpolating along both axes, each sample effectively covers its surrounding area, and by randomly sampling we cover more of the spatial sampling domain. Contrarily, sampling on a grid leaves large areas uncovered; we are effectively sampling at a much lower resolution than when using random sampling.

To test whether or not this improvement is intrinsically linked to the randomness of sampling, we also construct an alternative sampling method called shifted-grid sampling. Given a sampling grid with sensor distance $\Delta x$, shifted-grid sampling translates this grid a distance $\Delta a$ every frame, leading to an effective sample distance of $\Delta a \ll \Delta x$. This strategy, similarly as random sampling varies the sensor position over time, but does so in a deterministic and grid-based way. We show this sampling strategy in figure \ref{fig:burgers}a, while panels b and c confirm our hypothesis; shifted grid sampling (Blue) performs similarly to random sampling. Shifted-grid sampling relies on a densely sampled temporal axis 'compensating' for the sparsely sampled spatial axes. This makes off-grid sampling beneficial when either the time or space axis, but not both, can be sampled with a high resolution. In the experimental section we show that if both axes are sparsely sampled, we do not see a strong improvement over grid sampling. 

\subsection{Experimental data - 2D Advection-Diffusion}

\begin{figure*}[t]
    \centering
    \includegraphics[width = 0.95 \textwidth]{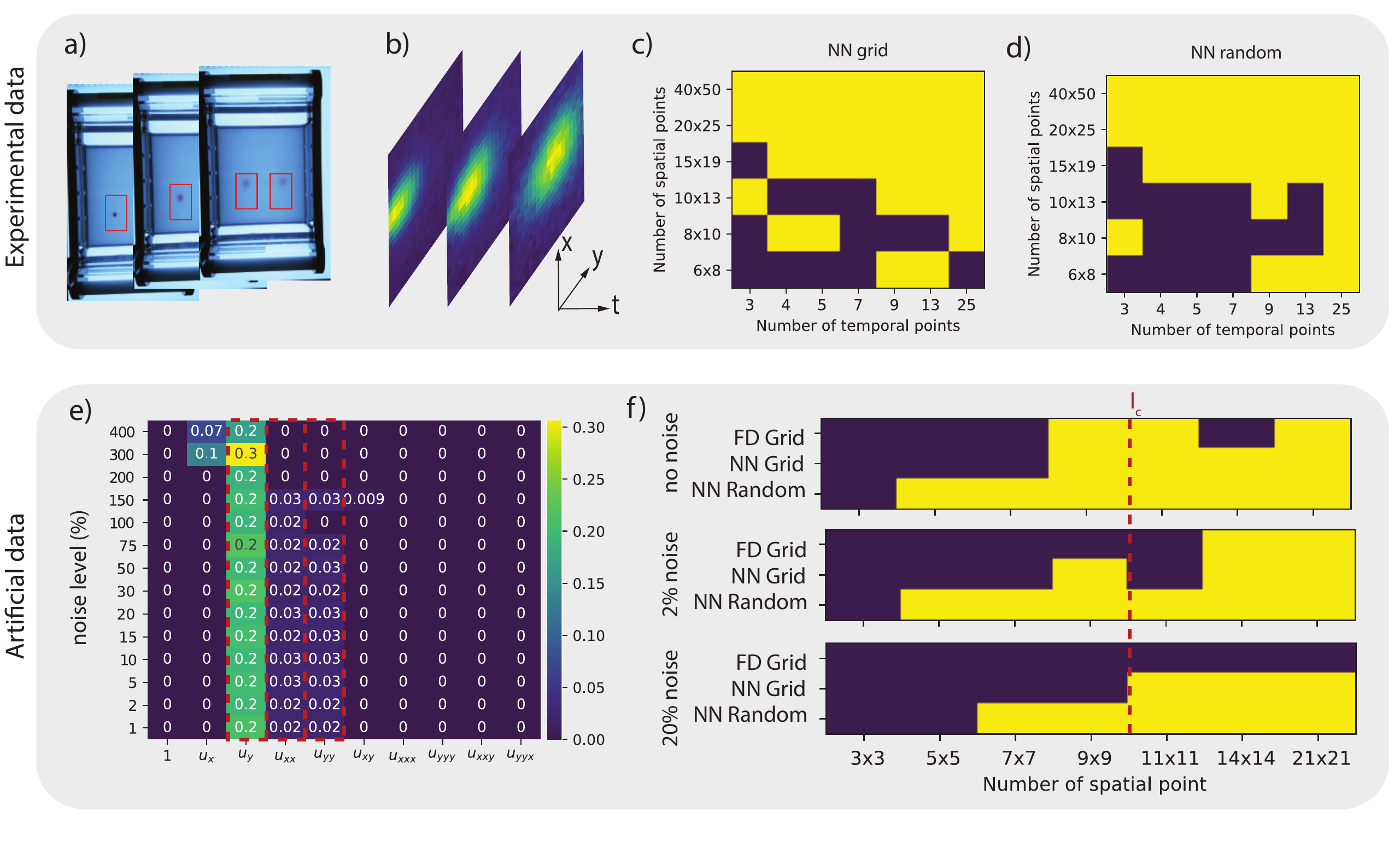}
    \caption{a) Experimental setup of the gel electrophoresis. b) Three time frames of the density  with a spatial resolution of 20x25. c) and d) Success diagram for the experimental data indicating correct model discovery (Yellow indicates the correct AD equation $u_t = D (u_{xx} + u_{yy}) + v u_y$ is discovered) as function of the spatial and temporal resolution for c) grid sampling and d) random sampling. e) Obtained mask and coefficients ($D = 0.025$ and and $v = (0, 0.2)$) for the artificial data-set as function of the noise level (11x11 spatial resolution). Hereby, yellow indicates the terms selected by the algorithm and the red dashed box the terms that are expected in the PDE. f) Success diagrams for various levels of additive noise, comparing the result of DeepMoD with a grid and random sampling strategy and the classical LassoCV algorithm on a Finite Difference (FD)-based library (after SVD filtering of the different frames).}
    \label{figure:AD}
\end{figure*}

In an electrophoresis experiment, a charged dye is pipetted in a gel over which a spatially uniform electric field is applied (see Figure \ref{figure:AD}a)). The dye passively diffuses in the gel and is advected by the applied electric field, giving rise to an advection-diffusion equation with advection in one direction: $u_t = D (u_{xx} + u_{yy}) + v u_y$. \cite{both2019} showed that \emph{DeepMoD} could discover the correct underlying equation from the full data-set (size 120 x 150 pixels and 25 frames). Here, we study how much we can sub-sample this data and still discover the advection-diffusion equation.

In figure \ref{figure:AD} c) and d) we perform grid based as well as a random sub-sampling of the data. The neural network-based method discovers the advection-diffusion equation on as few as 6 x 8 spatial sampling points with 13 time-points, or with 20 x 25 pixels on only 3 time-points. The minimum number of required samples is similar for grid and random sampling, confirming that when both axes are poorly sampled, there is no benefit to sample randomly. 

The smallest characteristic length scale in the problem is the width of the dye at $t=t_0$, which we estimate as $l_c \approx 10$ pixels. For the data presented in figure \ref{figure:AD}c) and \ref{figure:AD}d), at a resolution below $10 \times 13$ sensors classical approaches would inherently fail, even if no noise was present in the data set. This is indeed what we observe: using a finite difference-based library we were unable to recover the advection-diffusion equation, even after denoising with SVD (See Appendix A for details).


The use of a neural network and random sampling lead to non-deterministic behaviour: the neural network training dynamics depend on its initialization and two randomly sampled datasets of similar size might not lead to similar results. In practice this leads to a 'soft' decision boundary, where a fraction of a set of runs with different initialization and datasets fail. We discuss and study this issue in appendix B.

\paragraph{Noiseless synthetic data set} 

To further confirm our results from the previous section, we simulate the experiment by solving the 2D advection-diffusion with a Gaussian initial condition and experimentally determined parameters ($D = 0.025$ and and $v = (0, 0.2)$. Figure \ref{figure:AD}e) shows the selected terms and their magnitude as functions of the applied noise levels for a highly subsampled data-set (grid sampling, spatial resolution of 11x11 and temporal resolution 14). The correct AD equation is recovered up to noise levels of 100$\%$ (See figure \ref{figure:AD}e), confirming the noise robustness of the NN-based model discovery. In panel f) we compare the deep-learning based model discovery using grid and random sampling with classical methods for various noise levels and sensor spacing with a fixed temporal resolution of 81 frames (Data for the FD was pre-processed with a SVD filter, see SI for further details). We confirm that, similarly to the Burgers example of the previous section, the correct underlying PDE is discovered even below the smallest characteristic length-scale in the problem (indicated by a red dashed line in figure \ref{figure:AD}f).

This figure confirms our three main conclusions: 1) In the noiseless limit, classical approaches are only slightly less performing than NN-based model discovery for grid sampling. 2) Increasing the noise level dramatically impacts classical model discovery while barely impacting NN-based model discovery and 3) random sampling over space considerably improves performance, allowing model discovery with roughly 4-8 times fewer sample points for this particular data-set (depending on the noise level). 

\subsection{Experimental data - Cable equation}
\begin{figure}
\centering
    \includegraphics[width = 0.95 \textwidth]{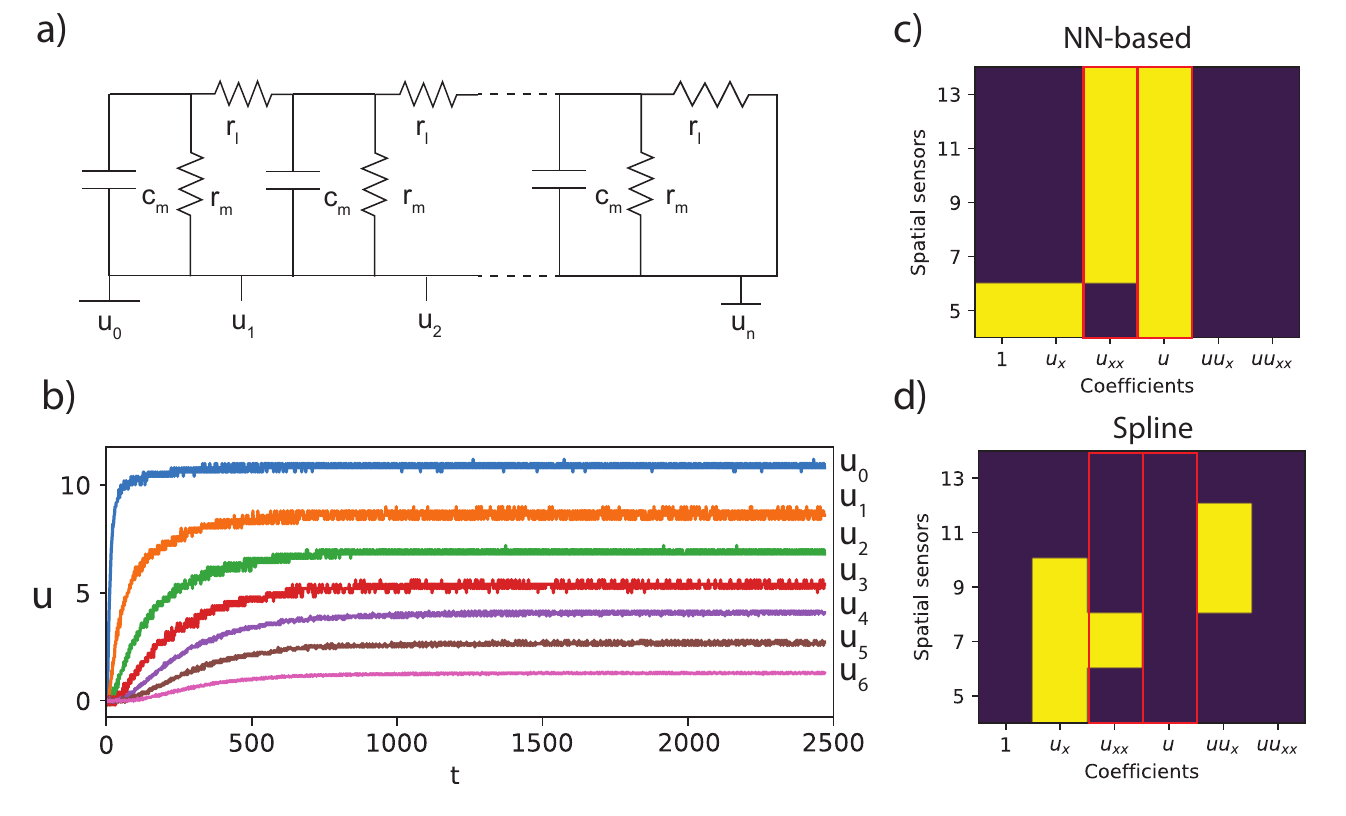}
    \caption{a) Schematic overview of the electronic setup to generate the cable equation. b) The voltage drop $u$ as function of time for various positions along the circuit for a circuit with 7 elements. The mask obtained for c) NN-based and d) cross validated Lasso with spline based library model discovery (Yellow indicates the term was recovered by the algorithm). The red boxes indicate the two expected terms in the equation.}
    \label{figure:CE}
\end{figure}

 Applying a constant voltage to a RC-circuit with longitudinal resistance (see figure \ref{figure:CE} a) result in time-dependent voltage increase throughout the circuit due to the charging of the capacitors. This rise is modeled by the cable equation, which is essentially a reaction-diffusion equation $ u_t =  u_{xx}/ (R_l C) + u/(R_m C)$ with $C$ the capacitance, $R_l$ the longitudinal resistance and $R_m$ the parallel resistance of the circuit. The discrete nature of the experiment automatically gives $\Delta x = O(l_c)$. We consider an extreme case where we only have seven sensors throughout the circuit (i.e. spatial axis), but take 2500 samples along the time axis. Figure \ref{figure:CE}b shows the measured voltage at these seven elements. Initially, all the capacitors are uncharged and we observe a sharp voltage increase at the first element. As the capacitors charge, this further propagates through the circuit, charging the capacitors and resulting in the curves shown in the figure. We apply both a classical approach with the library generated with splines and DeepMoD to a varying amount of elements. Figure \ref{figure:CE} c and d show that the DeepMoD discovers the cable equation with as few as seven elements, whereas classical methods are unable to find the cable equation at any number of elements. 

\section{Discussion and future work}
In this paper we showed how a deep learning approach allows to discover partial differential equations from coarsely and off-grid sampled observations in time and space. The correct equations are discovered, even when the sensor spacing is larger than some data set’s characteristic length scale- an inaccessible regime when using numerical differentiation procedures. We have also shown that the presence of noise quickly deteriorates the performance of classical methods, whereas the neural network based method is much less affected. However, in the limit of very sparse data, model discovery can be sensitive to the exact positioning of the sensors, hence sensitive to where exactly on the grid the samples are drawn. Future work could investigate the upper limit of the characteristic length scale above which the approach consistently starts failing and how it relates to the spectrum of the data. We will also focus in the future on including more structure in the interpolation for better convergence and initialization robustness, for example using Gaussian Processes.

\section*{Acknowledgments}
This work received support from the CRI Research Fellowship. We thank the Bettencourt Schueller Foundation long term partnership and NVidia for supplying the GPU under the Academic Grant program. We would also like to thank the authors and contributors of Numpy (\cite{harris_array_2020}), Scipy (\cite{virtanen_scipy_2020}), Scikit-learn (\cite{pedregosa2011scikit}), Matplotlib (\cite{hunter_matplotlib_2007}), Ipython (\cite{perez_ipython_2007}), and Pytorch (\cite{paszke_pytorch_2019}) for making our work possible through their open-source software. The authors declare no competing interest.

\bibliography{references}

\begin{thebibliography}{39}
\providecommand{\natexlab}[1]{#1}
\providecommand{\url}[1]{\texttt{#1}}
\expandafter\ifx\csname urlstyle\endcsname\relax
  \providecommand{\doi}[1]{doi: #1}\else
  \providecommand{\doi}{doi: \begingroup \urlstyle{rm}\Url}\fi

\bibitem[Baydin et~al.(2017)Baydin, Pearlmutter, Radul, and
  Siskind]{baydin2017automatic}
At{\i}l{\i}m~G{\"u}nes Baydin, Barak~A Pearlmutter, Alexey~Andreyevich Radul,
  and Jeffrey~Mark Siskind.
\newblock Automatic differentiation in machine learning: a survey.
\newblock \emph{The Journal of Machine Learning Research}, 18\penalty0
  (1):\penalty0 5595--5637, 2017.

\bibitem[Berg \& Nyström(2019)Berg and Nyström]{berg_data-driven_2019}
Jens Berg and Kaj Nyström.
\newblock Data-driven discovery of {PDEs} in complex datasets.
\newblock \emph{Journal of Computational Physics}, 384:\penalty0 239--252, May
  2019.
\newblock ISSN 00219991.
\newblock \doi{10.1016/j.jcp.2019.01.036}.
\newblock URL \url{http://arxiv.org/abs/1808.10788}.
\newblock arXiv: 1808.10788.

\bibitem[Bolton \& Zanna(2019)Bolton and Zanna]{bolton2019applications}
Thomas Bolton and Laure Zanna.
\newblock Applications of deep learning to ocean data inference and subgrid
  parameterization.
\newblock \emph{Journal of Advances in Modeling Earth Systems}, 11\penalty0
  (1):\penalty0 376--399, 2019.

\bibitem[Both \& Kusters(2020)Both and Kusters]{both2020}
Gert-Jan Both and Remy Kusters.
\newblock Sparsely constrained neural networks for model discovery of pdes.
\newblock \emph{arXiv preprint arXiv:2011.04336}, 2020.

\bibitem[Both et~al.(2021)Both, Choudhury, Sens, and Kusters]{both2019}
Gert-Jan Both, Subham Choudhury, Pierre Sens, and Remy Kusters.
\newblock Deepmod: Deep learning for model discovery in noisy data.
\newblock \emph{Journal of Computational Physics}, 428:\penalty0 109985, 2021.

\bibitem[Brunton et~al.(2013)Brunton, Brunton, Proctor, and Kutz]{brunton2013}
Bingni~W Brunton, Steven~L Brunton, Joshua~L Proctor, and J~Nathan Kutz.
\newblock Optimal sensor placement and enhanced sparsity for classification.
\newblock \emph{arXiv preprint arXiv:1310.4217}, 2013.

\bibitem[Brunton et~al.(2015)Brunton, Proctor, Tu, and Kutz]{brunton2015}
Steven~L Brunton, Joshua~L Proctor, Jonathan~H Tu, and J~Nathan Kutz.
\newblock Compressed sensing and dynamic mode decomposition.
\newblock \emph{Journal of computational dynamics}, 2\penalty0 (2):\penalty0
  165, 2015.

\bibitem[Brunton et~al.(2016)Brunton, Proctor, and
  Kutz]{brunton_discovering_2016}
Steven~L. Brunton, Joshua~L. Proctor, and J.~Nathan Kutz.
\newblock Discovering governing equations from data by sparse identification of
  nonlinear dynamical systems.
\newblock \emph{Proceedings of the National Academy of Sciences}, 113\penalty0
  (15):\penalty0 3932--3937, April 2016.
\newblock ISSN 0027-8424, 1091-6490.
\newblock \doi{10.1073/pnas.1517384113}.
\newblock URL \url{http://www.pnas.org/lookup/doi/10.1073/pnas.1517384113}.

\bibitem[Champion et~al.(2019{\natexlab{a}})Champion, Lusch, Kutz, and
  Brunton]{champion_data-driven_2019}
Kathleen Champion, Bethany Lusch, J.~Nathan Kutz, and Steven~L. Brunton.
\newblock Data-driven discovery of coordinates and governing equations.
\newblock \emph{arXiv:1904.02107 [stat]}, March 2019{\natexlab{a}}.
\newblock URL \url{http://arxiv.org/abs/1904.02107}.
\newblock arXiv: 1904.02107.

\bibitem[Champion et~al.(2019{\natexlab{b}})Champion, Brunton, and
  Kutz]{champion2019}
Kathleen~P Champion, Steven~L Brunton, and J~Nathan Kutz.
\newblock Discovery of nonlinear multiscale systems: Sampling strategies and
  embeddings.
\newblock \emph{SIAM Journal on Applied Dynamical Systems}, 18\penalty0
  (1):\penalty0 312--333, 2019{\natexlab{b}}.

\bibitem[Chen et~al.(2020)Chen, Liu, and Sun]{chen_deep_2020}
Zhao Chen, Yang Liu, and Hao Sun.
\newblock Deep learning of physical laws from scarce data.
\newblock \emph{arXiv:2005.03448 [physics, stat]}, May 2020.
\newblock URL \url{http://arxiv.org/abs/2005.03448}.
\newblock arXiv: 2005.03448.

\bibitem[Cranmer et~al.(2020)Cranmer, Greydanus, Hoyer, Battaglia, Spergel, and
  Ho]{cranmer_lagrangian_2020}
Miles Cranmer, Sam Greydanus, Stephan Hoyer, Peter Battaglia, David Spergel,
  and Shirley Ho.
\newblock Lagrangian {Neural} {Networks}.
\newblock \emph{arXiv:2003.04630 [physics, stat]}, March 2020.
\newblock URL \url{http://arxiv.org/abs/2003.04630}.
\newblock arXiv: 2003.04630.

\bibitem[Greydanus et~al.(2019)Greydanus, Dzamba, and
  Yosinski]{greydanus_hamiltonian_2019}
Sam Greydanus, Misko Dzamba, and Jason Yosinski.
\newblock Hamiltonian {Neural} {Networks}.
\newblock \emph{arXiv:1906.01563 [cs]}, September 2019.
\newblock URL \url{http://arxiv.org/abs/1906.01563}.
\newblock arXiv: 1906.01563.

\bibitem[Harris et~al.(2020)Harris, Millman, van~der Walt, Gommers, Virtanen,
  Cournapeau, Wieser, Taylor, Berg, Smith, Kern, Picus, Hoyer, van Kerkwijk,
  Brett, Haldane, del Río, Wiebe, Peterson, Gérard-Marchant, Sheppard, Reddy,
  Weckesser, Abbasi, Gohlke, and Oliphant]{harris_array_2020}
Charles~R. Harris, K.~Jarrod Millman, Stéfan~J. van~der Walt, Ralf Gommers,
  Pauli Virtanen, David Cournapeau, Eric Wieser, Julian Taylor, Sebastian Berg,
  Nathaniel~J. Smith, Robert Kern, Matti Picus, Stephan Hoyer, Marten~H. van
  Kerkwijk, Matthew Brett, Allan Haldane, Jaime~Fernández del Río, Mark
  Wiebe, Pearu Peterson, Pierre Gérard-Marchant, Kevin Sheppard, Tyler Reddy,
  Warren Weckesser, Hameer Abbasi, Christoph Gohlke, and Travis~E. Oliphant.
\newblock Array programming with {NumPy}.
\newblock \emph{Nature}, 585\penalty0 (7825):\penalty0 357--362, September
  2020.
\newblock ISSN 0028-0836, 1476-4687.
\newblock \doi{10.1038/s41586-020-2649-2}.
\newblock URL \url{http://www.nature.com/articles/s41586-020-2649-2}.

\bibitem[Hunter(2007)]{hunter_matplotlib_2007}
John~D. Hunter.
\newblock Matplotlib: {A} {2D} {Graphics} {Environment}.
\newblock \emph{Computing in Science Engineering}, 9\penalty0 (3):\penalty0
  90--95, May 2007.
\newblock ISSN 1558-366X.
\newblock \doi{10.1109/MCSE.2007.55}.
\newblock Conference Name: Computing in Science Engineering.

\bibitem[Loiseau et~al.(2017)Loiseau, Noack, and Brunton]{loiseau2017sparse}
Jean-Christophe Loiseau, Bernd~R Noack, and Steven~L Brunton.
\newblock Sparse reduced-order modeling: sensor-based dynamics to full-state
  estimation.
\newblock \emph{arXiv preprint arXiv:1706.03531}, 2017.

\bibitem[Long et~al.(2018)Long, Lu, Ma, and Dong]{long2018}
Zichao Long, Yiping Lu, Xianzhong Ma, and Bin Dong.
\newblock Pde-net: Learning pdes from data.
\newblock In \emph{International Conference on Machine Learning}, pp.\
  3208--3216, 2018.

\bibitem[Maddu et~al.(2019)Maddu, Cheeseman, Sbalzarini, and
  Müller]{maddu_stability_2019}
Suryanarayana Maddu, Bevan~L. Cheeseman, Ivo~F. Sbalzarini, and Christian~L.
  Müller.
\newblock Stability selection enables robust learning of partial differential
  equations from limited noisy data.
\newblock \emph{arXiv:1907.07810 [physics]}, July 2019.
\newblock URL \url{http://arxiv.org/abs/1907.07810}.
\newblock arXiv: 1907.07810.

\bibitem[Mangan et~al.(2017)Mangan, Kutz, Brunton, and Proctor]{mangan2017}
Niall~M Mangan, J~Nathan Kutz, Steven~L Brunton, and Joshua~L Proctor.
\newblock Model selection for dynamical systems via sparse regression and
  information criteria.
\newblock \emph{Proceedings of the Royal Society A: Mathematical, Physical and
  Engineering Sciences}, 473\penalty0 (2204):\penalty0 20170009, 2017.

\bibitem[Manohar et~al.(2018)Manohar, Brunton, Kutz, and Brunton]{manohar2018}
Krithika Manohar, Bingni~W Brunton, J~Nathan Kutz, and Steven~L Brunton.
\newblock Data-driven sparse sensor placement for reconstruction: Demonstrating
  the benefits of exploiting known patterns.
\newblock \emph{IEEE Control Systems Magazine}, 38\penalty0 (3):\penalty0
  63--86, 2018.

\bibitem[Messenger \& Bortz(2020)Messenger and Bortz]{messenger2020}
Daniel~A Messenger and David~M Bortz.
\newblock Weak sindy for partial differential equations.
\newblock \emph{arXiv preprint arXiv:2007.02848}, 2020.

\bibitem[Paszke et~al.(2019)Paszke, Gross, Massa, Lerer, Bradbury, Chanan,
  Killeen, Lin, Gimelshein, Antiga, Desmaison, Köpf, Yang, DeVito, Raison,
  Tejani, Chilamkurthy, Steiner, Fang, Bai, and Chintala]{paszke_pytorch_2019}
Adam Paszke, Sam Gross, Francisco Massa, Adam Lerer, James Bradbury, Gregory
  Chanan, Trevor Killeen, Zeming Lin, Natalia Gimelshein, Luca Antiga, Alban
  Desmaison, Andreas Köpf, Edward Yang, Zach DeVito, Martin Raison, Alykhan
  Tejani, Sasank Chilamkurthy, Benoit Steiner, Lu~Fang, Junjie Bai, and Soumith
  Chintala.
\newblock {PyTorch}: {An} {Imperative} {Style}, {High}-{Performance} {Deep}
  {Learning} {Library}.
\newblock \emph{arXiv:1912.01703 [cs, stat]}, December 2019.
\newblock URL \url{http://arxiv.org/abs/1912.01703}.
\newblock arXiv: 1912.01703.

\bibitem[Pedregosa et~al.(2011)Pedregosa, Varoquaux, Gramfort, Michel, Thirion,
  Grisel, Blondel, Prettenhofer, Weiss, Dubourg, et~al.]{pedregosa2011scikit}
Fabian Pedregosa, Ga{\"e}l Varoquaux, Alexandre Gramfort, Vincent Michel,
  Bertrand Thirion, Olivier Grisel, Mathieu Blondel, Peter Prettenhofer, Ron
  Weiss, Vincent Dubourg, et~al.
\newblock Scikit-learn: Machine learning in python.
\newblock \emph{the Journal of machine Learning research}, 12:\penalty0
  2825--2830, 2011.

\bibitem[Perez \& Granger(2007)Perez and Granger]{perez_ipython_2007}
F.~Perez and B.~E. Granger.
\newblock {IPython}: {A} {System} for {Interactive} {Scientific} {Computing}.
\newblock \emph{Computing in Science Engineering}, 9\penalty0 (3):\penalty0
  21--29, May 2007.
\newblock ISSN 1558-366X.
\newblock \doi{10.1109/MCSE.2007.53}.

\bibitem[Rackauckas et~al.(2020)Rackauckas, Ma, Martensen, Warner, Zubov,
  Supekar, Skinner, and Ramadhan]{rackauckas_universal_2020}
Christopher Rackauckas, Yingbo Ma, Julius Martensen, Collin Warner, Kirill
  Zubov, Rohit Supekar, Dominic Skinner, and Ali Ramadhan.
\newblock Universal {Differential} {Equations} for {Scientific} {Machine}
  {Learning}.
\newblock \emph{arXiv:2001.04385 [cs, math, q-bio, stat]}, January 2020.
\newblock URL \url{http://arxiv.org/abs/2001.04385}.
\newblock arXiv: 2001.04385.

\bibitem[Raissi et~al.(2017{\natexlab{a}})Raissi, Perdikaris, and
  Karniadakis]{raissi_physics_2017}
Maziar Raissi, Paris Perdikaris, and George~Em Karniadakis.
\newblock Physics {Informed} {Deep} {Learning} ({Part} {I}): {Data}-driven
  {Solutions} of {Nonlinear} {Partial} {Differential} {Equations}.
\newblock \emph{arXiv:1711.10561 [cs, math, stat]}, November
  2017{\natexlab{a}}.
\newblock URL \url{http://arxiv.org/abs/1711.10561}.
\newblock arXiv: 1711.10561.

\bibitem[Raissi et~al.(2017{\natexlab{b}})Raissi, Perdikaris, and
  Karniadakis]{raissi_physics_2017-1}
Maziar Raissi, Paris Perdikaris, and George~Em Karniadakis.
\newblock Physics {Informed} {Deep} {Learning} ({Part} {II}): {Data}-driven
  {Discovery} of {Nonlinear} {Partial} {Differential} {Equations}.
\newblock \emph{arXiv:1711.10566 [cs, math, stat]}, November
  2017{\natexlab{b}}.
\newblock URL \url{http://arxiv.org/abs/1711.10566}.
\newblock arXiv: 1711.10566.

\bibitem[Rudy et~al.(2017{\natexlab{a}})Rudy, Brunton, Proctor, and
  Kutz]{rudy2017}
Samuel~H Rudy, Steven~L Brunton, Joshua~L Proctor, and J~Nathan Kutz.
\newblock Data-driven discovery of partial differential equations.
\newblock \emph{Science Advances}, 3\penalty0 (4):\penalty0 e1602614,
  2017{\natexlab{a}}.

\bibitem[Rudy et~al.(2017{\natexlab{b}})Rudy, Brunton, Proctor, and
  Kutz]{rudy_data-driven_2017}
Samuel~H. Rudy, Steven~L. Brunton, Joshua~L. Proctor, and J.~Nathan Kutz.
\newblock Data-driven discovery of partial differential equations.
\newblock \emph{Science Advances}, 3\penalty0 (4):\penalty0 e1602614, April
  2017{\natexlab{b}}.
\newblock ISSN 2375-2548.
\newblock \doi{10.1126/sciadv.1602614}.
\newblock URL
  \url{http://advances.sciencemag.org/lookup/doi/10.1126/sciadv.1602614}.

\bibitem[Rudy et~al.(2019)Rudy, Kutz, and Brunton]{rudy_deep_2019}
Samuel~H. Rudy, J.~Nathan Kutz, and Steven~L. Brunton.
\newblock Deep learning of dynamics and signal-noise decomposition with
  time-stepping constraints.
\newblock \emph{Journal of Computational Physics}, 396:\penalty0 483--506,
  November 2019.
\newblock ISSN 00219991.
\newblock \doi{10.1016/j.jcp.2019.06.056}.
\newblock URL \url{http://arxiv.org/abs/1808.02578}.
\newblock arXiv: 1808.02578.

\bibitem[Sanchez-Gonzalez et~al.(2018)Sanchez-Gonzalez, Heess, Springenberg,
  Merel, Riedmiller, Hadsell, and Battaglia]{sanchez-gonzalez_graph_2018}
Alvaro Sanchez-Gonzalez, Nicolas Heess, Jost~Tobias Springenberg, Josh Merel,
  Martin Riedmiller, Raia Hadsell, and Peter Battaglia.
\newblock Graph networks as learnable physics engines for inference and
  control.
\newblock \emph{arXiv:1806.01242 [cs, stat]}, June 2018.
\newblock URL \url{http://arxiv.org/abs/1806.01242}.
\newblock arXiv: 1806.01242.

\bibitem[Sanchez-Pi et~al.(2020)Sanchez-Pi, Marti, Abreu, Bernard, de~Vargas,
  Eveillard, Maass, Marquet, Sainte-Marie, Salomon,
  et~al.]{sanchez2020artificial}
Nayat Sanchez-Pi, Luis Marti, Andr{\'e} Abreu, Olivier Bernard, Colomban
  de~Vargas, Damien Eveillard, Alejandro Maass, Pablo~A Marquet, Jacques
  Sainte-Marie, Julien Salomon, et~al.
\newblock Artificial intelligence, machine learning and modeling for
  understanding the oceans and climate change.
\newblock In \emph{NeurIPS 2020 Workshop-Tackling Climate Change with Machine
  Learning}, 2020.

\bibitem[Schaeffer(2017)]{schaeffer_learning_2017}
Hayden Schaeffer.
\newblock Learning partial differential equations via data discovery and sparse
  optimization.
\newblock \emph{Proceedings of the Royal Society A: Mathematical, Physical and
  Engineering Sciences}, 473\penalty0 (2197):\penalty0 20160446, January 2017.
\newblock ISSN 1364-5021, 1471-2946.
\newblock \doi{10.1098/rspa.2016.0446}.
\newblock URL
  \url{https://royalsocietypublishing.org/doi/10.1098/rspa.2016.0446}.

\bibitem[Schaeffer et~al.(2018)Schaeffer, Tran, and
  Ward]{schaeffer2018extracting}
Hayden Schaeffer, Giang Tran, and Rachel Ward.
\newblock Extracting sparse high-dimensional dynamics from limited data.
\newblock \emph{SIAM Journal on Applied Mathematics}, 78\penalty0 (6):\penalty0
  3279--3295, 2018.

\bibitem[Seo \& Liu(2019)Seo and Liu]{seo_differentiable_2019}
Sungyong Seo and Yan Liu.
\newblock Differentiable {Physics}-informed {Graph} {Networks}.
\newblock \emph{arXiv:1902.02950 [cs, stat]}, February 2019.
\newblock URL \url{http://arxiv.org/abs/1902.02950}.
\newblock arXiv: 1902.02950.

\bibitem[Vaddireddy et~al.(2020)Vaddireddy, Rasheed, Staples, and
  San]{vaddireddy2020}
Harsha Vaddireddy, Adil Rasheed, Anne~E Staples, and Omer San.
\newblock Feature engineering and symbolic regression methods for detecting
  hidden physics from sparse sensor observation data.
\newblock \emph{Physics of Fluids}, 32\penalty0 (1):\penalty0 015113, 2020.

\bibitem[Virtanen et~al.(2020)Virtanen, Gommers, Oliphant, Haberland, Reddy,
  Cournapeau, Burovski, Peterson, Weckesser, Bright, van~der Walt, Brett,
  Wilson, Millman, Mayorov, Nelson, Jones, Kern, Larson, Carey, Polat, Feng,
  Moore, VanderPlas, Laxalde, Perktold, Cimrman, Henriksen, Quintero, Harris,
  Archibald, Ribeiro, Pedregosa, van Mulbregt, and
  Contributors]{virtanen_scipy_2020}
Pauli Virtanen, Ralf Gommers, Travis~E. Oliphant, Matt Haberland, Tyler Reddy,
  David Cournapeau, Evgeni Burovski, Pearu Peterson, Warren Weckesser, Jonathan
  Bright, Stéfan~J. van~der Walt, Matthew Brett, Joshua Wilson, K.~Jarrod
  Millman, Nikolay Mayorov, Andrew R.~J. Nelson, Eric Jones, Robert Kern, Eric
  Larson, C.~J. Carey, İlhan Polat, Yu~Feng, Eric~W. Moore, Jake VanderPlas,
  Denis Laxalde, Josef Perktold, Robert Cimrman, Ian Henriksen, E.~A. Quintero,
  Charles~R. Harris, Anne~M. Archibald, Antônio~H. Ribeiro, Fabian Pedregosa,
  Paul van Mulbregt, and SciPy 1~0 Contributors.
\newblock {SciPy} 1.0--{Fundamental} {Algorithms} for {Scientific} {Computing}
  in {Python}.
\newblock \emph{Nature Methods}, 17\penalty0 (3):\penalty0 261--272, March
  2020.
\newblock ISSN 1548-7091, 1548-7105.
\newblock \doi{10.1038/s41592-019-0686-2}.
\newblock URL \url{http://arxiv.org/abs/1907.10121}.
\newblock arXiv: 1907.10121.

\bibitem[Wang et~al.(2019)Wang, Li, and Chen]{wang2019}
Zhi Wang, Han-Xiong Li, and Chunlin Chen.
\newblock Reinforcement learning-based optimal sensor placement for
  spatiotemporal modeling.
\newblock \emph{IEEE transactions on cybernetics}, 50\penalty0 (6):\penalty0
  2861--2871, 2019.

\bibitem[Yuan et~al.(2019)Yuan, Li, Li, Jiang, Tang, Zhang, Liu, Goncalves,
  Voss, Li, Kurths, and Ding]{yuan_machine_2019}
Ye~Yuan, Junlin Li, Liang Li, Frank Jiang, Xiuchuan Tang, Fumin Zhang, Sheng
  Liu, Jorge Goncalves, Henning~U. Voss, Xiuting Li, Jürgen Kurths, and Han
  Ding.
\newblock Machine {Discovery} of {Partial} {Differential} {Equations} from
  {Spatiotemporal} {Data}.
\newblock \emph{arXiv:1909.06730 [physics, stat]}, September 2019.
\newblock URL \url{http://arxiv.org/abs/1909.06730}.
\newblock arXiv: 1909.06730.

\end{thebibliography}
\bibliographystyle{iclr2021_conference}

\appendix

\section{Reproducibility}

\subsection{Hyperparameters}

\paragraph{DeepMoD:} In this paper we use the neural network based model discovery tool DeepMoD \footnote{github.com/PhIMaL/DeePyMoD}. Every experiment uses a neural network with tanh-activation functions and 4 layers of 30 neurons with random initialization, and an Adam optimizer with default learning rate $10^{-3}$ and $\beta = (0.9,0.9)$. The sparsity scheduler has a patience of 500 epochs and a periodicity of 50 epochs \cite{both2020}.We use a cross validated, thresholded Lasso sparsity selection with a threshold of 0.2 and otherwise default parameters from the Sklearn implementation \cite{pedregosa2011scikit}

\paragraph{Spline interpolation:} For fitting the spline interpolation in both the Burgers as well as the Cable equation, we use a smoothing parameter of $s = 0.01$ in the case of noisy data and 5th order splines.

\paragraph{Finite difference and SVD filter:} To construct the function library of the 2D Advection diffusion equation we use a second-order accurate central difference scheme. For the 2D advection-diffusion data, the data was denoised using by decomposing it using the SVD and (\cite{harris_array_2020}) selecting the 3 largest modes from the signal.

\paragraph{Noise on synthetic data:} We add white noise to the data with a strength relative to the standard deviation of the data, i.e. $50\%$ noise corresponds to $0.5 \cdot \sigma$.

\subsection{Data preparation}

\paragraph{Burgers equation:} Using the Cole-Hopf transform, the Burgers equation described in the main text reduces to the heat equation and can be solved exactly for a delta peak initial condition to give, 

\begin{equation}
    u(x, t) = \sqrt{\frac{\nu}{t\pi}}\left(\frac{(e^{R}-1)e^{-z^2}}{1 + \frac{\left(e^{R}-1\right)}{2}\text{erfc}(z)}\right).
    \label{eq:burgers_dimless}
\end{equation}

where $R = A / 2 \nu$ and $z = x / \sqrt{4\nu t}$, a dimensionless coordinate. The characteristic length-scale is thus the smallest on in the system; for our case study $l_c = \sqrt{ 4 \nu t}|_{t=t_0}$. We use a function library containing all combinations of up to third order spatial derivative and second order polynomials in $u$, for a total of 12 terms, i.e.,
\begin{equation}
   \Theta = \left[1,u_x,u_{xx},u_{xxx}, u, uu_x, uu_{xx}, uu_{xxx},u^2,u^2u_x, u^2u_{xx}, u^2u_{xxx}\right].
\end{equation}

\paragraph{Cable equation:} We measured the passive voltage drop across a RC-circuit coupled to a longitudinal resistance (See Fig. \ref{figure:CE}A). This voltage drop across the circuit typically serves to model the passive voltage transmission through a dendrite, and is described by the so-called cable equation,
\begin{equation}
    u_t = \frac{1}{R_l C} u_{xx} - \frac{1}{R_m C} u.
\end{equation}
Here $C$ is the capacitance. $R_l$ the longitudinal resistance and $R_m$ the membrane resistance. This equation can be discretizised by an electric circuit, consisting of a serial set of $n$ longitudinal resistors, $r_i$, membrane resistors, $r_m$, and capacitors, $c_m$. Using Ohm’s ans Kirchhoff’s law, the discretized versioan of an array of these elements read, 
\begin{equation}
\frac{du_{i}}{dt}= \frac{(u_{i-1}+2 u_i + u_{i+1})}{c_{m}r_{l}} - \frac{u_i}{c_{m}r_{m}}.
\end{equation}

We use a breadboard using structures imitating GMMs, using only standard electronics hardware ($r_m = 10 k \Omega$, $r_c = 270 \Omega$ and $c_m = 680 mF$). We applied a voltage profile across the electronics structure using an arbitrary wave generator (AWG) (\textit{mhinstek MHS-2300A})  and used a dual channel oscilloscope (\textit{Voltcraft DSO-1062D}) to measure the voltage at various positions along the circuitry. These positions along the circuitry are the the spatial dimension of the cable equation. We varied the amount of elements between 5 and 13, mimicking various spatial discretizations. At every sensor, we collected 2500 data-points. We trained the model discovery algorithm on a function library up to first order polynomials and second order derivatives.

\subsection{2D Advection Diffusion}

\paragraph{Experiment:} We consider the 2D advection-diffusion process described by,
\begin{equation}
  u_t = -\nabla\cdot\left(-D\nabla u + \vec{v} u \right). 
\end{equation}

Here $\vec{v}$ is the velocity vector describing the advection and $D$ is the diffusion coefficient. We measure a time-series of images from an electrophoresis experiment, tracking the advection-diffusion of a charged purple loading dye under the influence of a spatially uniform electric field. We capture a set of 25 images with a resolution of 120x150 pixels and show the resultant 2D density field for three separate time-frames (in arbitrary units) in Fig. \ref{figure:AD}a, by subtracting the reference image (no dye present). The dye displays a diffusive and advective motion with constant velocity $\vec{v}$, which is related to the strength of the applied electric field. We use this data-set to asses the impact of temporal as well as spatial sensor density on the model discovery task.  We used a cross validated thresholded Lasso sparsity selection with a threshold of 0.05 and a function library containing all combinations of up to third order spatial derivative and second order polynomials in $u$, for a total of 10 terms, 
\begin{equation}
    {\Theta} = \left[1,u_x,u_y, u_{xx},u_{yy}, u_{xy}, u_{xxx}, u_{yyy}, u_{xxy}, u_{xyy}\right].
\end{equation}
%

\section{Sensitivity to the random sampling}

In this Appendix we discuss the sensitivity of the deep learning based approach, DeepMoD, w.r.t. the set of random samples selected, in particular is the limit $\Delta x /l_c > 1$. In order to show the impact of the random set of samples drawn we perform 10 runs of DeepMoD with otherwise identical parameters (1000 samples drawn and $10\%$ white noise and otherwise identical parameters as discussed in Appendix A). In Fig. \ref{figure:app}a we show the outcome for $\Delta x/l_c = 2 > 1$ indicated that in 7 of the 10 cases the correct equation is discovered and in 3 of the 10 cases this is not the case. In Fig. \ref{figure:app}b) we repeat this as function of $\Delta x/l_c$ and show that the larger the average distance between the samples becomes, the more pronounced the discrepancy between discovered models becomes. We have also tested the impact of the initialization of the neural network on the outcome, with identical set of samples and parameters, but this had little impact to the obtained PDE.

\begin{figure*}[t]
    \centering
    \includegraphics[width=0.95 \textwidth]{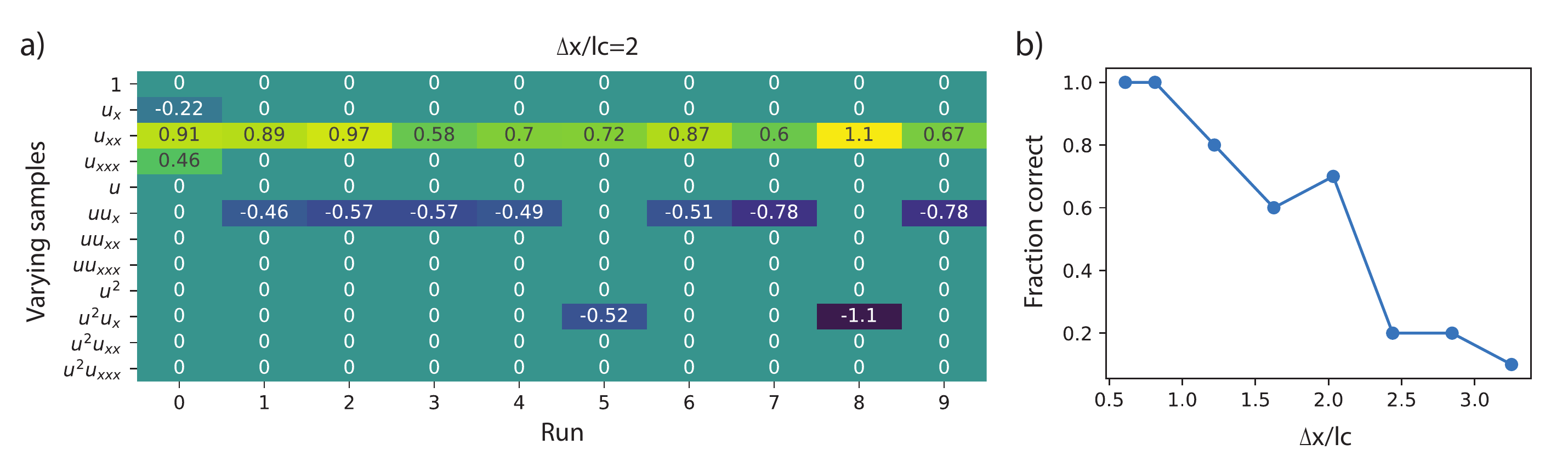}
    \caption{a) Coefficients obtained for the Burgers equation with 10$\%$ white noise for 10 separate runs with 10 sets of randomly sampled data-sets. b) Fraction of correctly discovered equations over 10 runs (with 10$\%$ white noise and 1000 samples per run) as function of the average distance between the samples, $\Delta x$, relative the the smallest characteristic length-scale $l_c$.}
    \label{figure:app}
\end{figure*}
\end{document}